\documentclass[aps,longbibliography,superscriptaddress,twocolumn]{revtex4-2}
\usepackage{graphicx}
\usepackage{stmaryrd}
\usepackage{amsmath, amsfonts,amssymb}
\usepackage{mathrsfs}
\usepackage{braket}
\usepackage{color}
\usepackage{xspace}
\usepackage{amscd}
\usepackage{comment}
\usepackage{bm}
\usepackage{bbm}
\definecolor{lightblue}{rgb}{0.13, 0.26, 0.99}

\usepackage[
colorlinks=true,
urlcolor=blue,
citecolor=blue,
linkcolor=blue,
hyperfootnotes=false]{hyperref}

\allowdisplaybreaks

\begin{document}

\title{Gapless chirality liquid with symmetry-protected edge spins}
\author{Shunsuke C. Furuya}
\affiliation{Department of Basic Science, University of Tokyo, Meguro, Tokyo 153-8902, Japan}
\author{Katsuhiro Morita}
\affiliation{Department of Physics, Faculty of Science and Technology,
Tokyo University of Science, Chiba 278-8510, Japan}

\date{\today}
\begin{abstract}
We report that a spin-1/2 tetrahedral Heisenberg chain realizes a gapless symmetry-protected topological (gSPT) phase characterized by the coexistence of the Tomonaga-Luttinger-liquid criticality due to chirality degrees of freedom and the symmetry-protected edge state due to spin degrees of freedom.
This gSPT phase has an interesting feature that no symmetry forbids the trivial spin gap opening but a discrete symmetry, $\mathbb Z_3\times\mathbb Z_2^T$, forbids the unique gapped ground state.
In the first part of the paper, we numerically show the coexistence of a critical entanglement entropy and a nontrivially degenerate entanglement spectrum based on the density-matrix renormalization group (DMRG) method.
Next, we clarify that chirality degrees of freedom form the Tomonaga-Luttinger liquid while spin degrees of freedom form the spin-1 Haldane state based on a degenerate perturbation theory.
Last but not least, we discuss the Lieb-Schultz-Mattis-type ingappability in the gSPT phase, using a local $\mathbb{Z}_3$ rotation.
We can thus characterize our gSPT phase as a symmetry-protected critical phase protected by the $\mathbb{Z}_3$ on-site symmetry, the $\mathbb{Z}_2^T$ time-reversal symmetry, the lattice translation symmetry, and the U(1) spin-rotation symmetry.
\end{abstract}

\maketitle

\section{Introduction}\label{sec:intro}

Topological phases of matter now became a major research subject of condensed matter physics.
Historically, researches on topological phases initially covered free fermion systems~\cite{TKNN1982,KaneMele2005_QSH,Schnyder2008_TI_TSC} and were later extended to strongly interacting systems.
Symmetry protected topological (SPT) phases are one of the best studied topological phases in strongly interacting quantum many-body systems~\cite{Chen2013_SPT,Gu2009_SPT}.
The (gapped) SPT phase is characterized as a quantum phase that has a unique gapped ground state with a short-range entanglement protected by symmetries.

Recently, gapless topological phases have drawn intensive attention~\cite{Matsuura2013_gapless,Scaffidi2017_gSPT,Hetenyi2020_gapless,Wang2020_gapless,Thorngren2021_gapless,He2021_gapless,Verresen2021_gapless,hidaka2022_gaplessSPT}.
Dirac and Weyl semimetals are their good examples in noninteracting or weakly interacting fermion systems~\cite{Matsuura2013_gapless}.
Naturally, people are giving their attention to gapless SPT (gSPT) phases in strongly interacting quantum many-body systems to get a deeper insight into topological phases of matter~\cite{Scaffidi2017_gSPT,Hetenyi2020_gapless,Wang2020_gapless,Thorngren2021_gapless,He2021_gapless,Verresen2021_gapless,hidaka2022_gaplessSPT}.
A naive definition of the gSPT phase is a gapless phase with a symmetry-protected entanglement.

However, we need careful considerations on the symmetry protection of the gSPT phase. 
The symmetry protection of the gSPT phase has two meanings:
the symmetry protection of the edge state and the symmetry protection of gapless low-energy states in bulk.
The symmetry protection in the former sense is nontrivial in the gSPT phase because of the absence of the excitation gap in bulk.
The bulk gap is a precondition for the symmetry protection of the gapped SPT phase.
The symmetry protection in the latter sense is also quite nontrivial.
This symmetry protection is defined as the prohibition of any trivial opening of the bulk gap~\cite{Furuya2017_spc}.
Here, we mean by trivial that the gap opens without any spontaneous symmetry breaking.
The impossibility of the trivial gap opening is also called ingappability~\cite{Hsieh2014_SPT_orientifold}.
The ingappability guarantees that the gSPT phase indeed qualifies as a phase of matter distinct from the other phases.

In this paper, we report a novel gSPT phase in geometrically frustrated quantum spin chains [Fig.~\ref{fig:tetrahedron}~(a)] and discuss its symmetry protection in both senses.
This gSPT phase hosts a critical chirality liquid with a symmetry-protected edge state.
The tetrahedral spin chain can trivially open a spin gap because the unit cell contains a tetrahedron of four localized spins [Fig.~\ref{fig:tetrahedron}~(a)].
Concerning the spin degrees of freedom, the ground state of this system looks like a valence-bond solid (VBS) state with a short-range entanglement.
This gSPT phase hosts gapless low-energy excitations originating from chirality degrees of freedom without interfering the VBS texture.

We organize this paper as follows.
Section~\ref{sec:model} defines our model.
We numerically investigate the model in Sec.~\ref{sec:num}, where we confirm that the model indeed has the gSPT phase from the entanglement point of view.
Section~\ref{sec:perturbation} discusses the gSPT phase with the degenerate perturbation theory.
The perturbation theory allows us to explicitly write down the low-energy effective Hamiltonian.
The effective Hamiltonian clarifies that the gSPT phase in our system is the chirality liquid with symmetry-protected edge spins.
Based on the results in Sec.~\ref{sec:perturbation}, we discuss the ingappability of the gSPT phase in Sec.~\ref{sec:LSM}.
Finally, we summarize the paper in Sec.~\ref{sec:summary}.

\section{Model}~\label{sec:model}

Our model has the following Hamiltonian under the periodic boundary condition (PBC):
\begin{align}
    \mathcal{H} 
    &=J\sum_{r=1}^L (\bm S_{r,1}\cdot \bm S_{r,2} + \bm S_{r,2}\cdot \bm S_{r,3} + \bm S_{r,3}\cdot \bm S_{r,1})
    \notag \\
    &\qquad +\alpha J\sum_{j=1}^L \bm{S}_{r,4}\cdot\bm{T}_{r}
    \notag \\
    &+J_t\sum_{r=1}^L\bm{S}_{r,4}\cdot\bm{T}_{r+1}+J_\ell\sum_{r=1}^L\sum_{n=1}^3\bm{S}_{r,n}\cdot\bm{S}_{r+1,n},
    \label{H_tetra}
\end{align}
where $\bm{S}_{r,n}$ is an $S=1/2$ localized spin at the $n$th vertex of the $r$th tetrahedron [Figs.~\ref{fig:tetrahedron}~(a) and (b)] and $\bm{T}_{r}=\sum_{n=1}^3\bm{S}_{r,n}$ is the total spin of the base triangle of the tetrahedron.
The tetrahedral spin chain \eqref{H_tetra} contains $4L$ spins, where we call $L$ a system length.

Three exchange couplings $J$, $J_t$, and $J_\ell$ are all positive (i.e., antiferromagnetic).
We limit the dimensionless parameter $\alpha$ to $\alpha < 0$ so that each tetrahedron has three antiferromagnetic $J$ bonds in the base triangle and three ferromagnetic $\alpha J$ bonds that bridge the base triangle with the top of the tetrahedron [Fig.~\ref{fig:tetrahedron}~(b)].
Throughout this paper, we assume $\max\{J_\ell, J_t\} \ll J$.

\section{Numerical results}\label{sec:num}

\begin{figure}[t!]
    \centering
    \includegraphics[width=\linewidth]{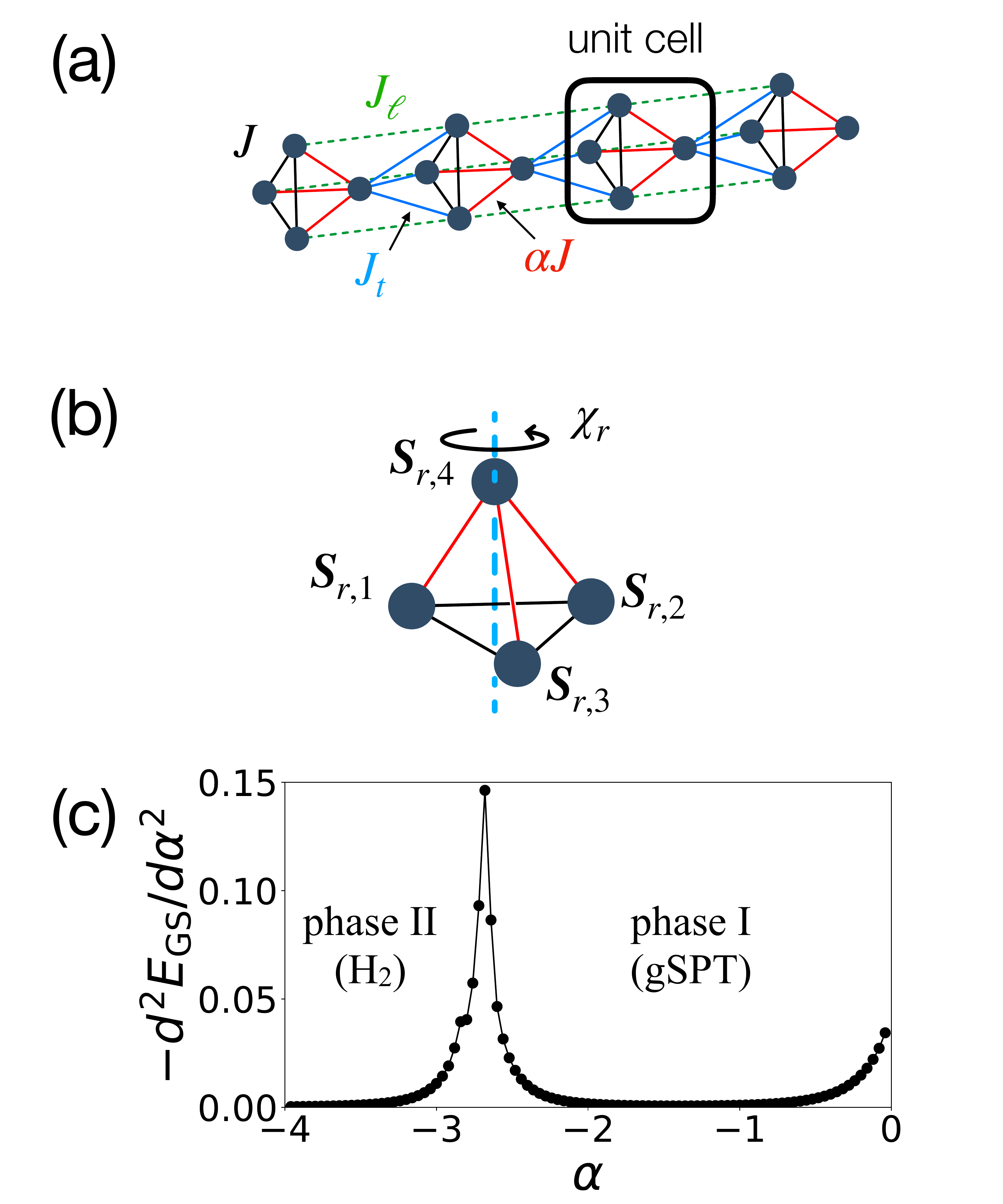}
    \caption{(a) Tetrahedral spin chain. The unit cell contains one tetrahedron with three antiferromagnetic $J>0$ and three ferromagnetic $\alpha J<0$ bonds.
    (b) Single tetrahedron with base triangle formed by $\bm S_{r,n}$ with $n=1,2,3$ and vertex $\bm S_{r,4}$. The solid curve around the dashed line piercing the base triangle and the vertex $\bm S_{r,4}$ depict the chirality $\chi_r = \pm 1$ of the tetrahedron.
    (c) Second derivative $-d^2E_{\mathrm{GS}}/d\alpha^2$ of GS energy calculated by using DMRG with parameters $L=30$, $J_t/J=0.2$, $J_\ell/J=0.1$.
    The derivative shows a singularity at $\alpha=\alpha_c\approx -0.27$.
    The point $\alpha=\alpha_c$ defines phase I and II.
    As we show later, we identify the phases I ($\alpha_c<\alpha<0$) and II ($\alpha<\alpha_c$) as the gSPT phase and the spin-2 Haldane phase (H$_2$), respectively.
    }
    \label{fig:tetrahedron}
\end{figure}

Let us numerically investigate the tetrahedral spin chain \eqref{H_tetra} to get insight into its quantum phases.
In this section, we fix $J_t/J=0.2$ and $J_\ell/J=0.1$ and vary $\alpha$ and the system length $L$.
Our calculations are based on the density-matrix renormalization group (DMRG) method with the open boundary condition (OBC).

The ground-state phase diagram $-\infty<\alpha<0$ contains two phases I and II separated by a quantum phase transition point $\alpha=\alpha_c$.
Let us start with locating the phase transition point.
Figure~\ref{fig:tetrahedron}~(c) shows a second derivative $-d^2E_{\text{GS}}/d\alpha^2$ of the GS energy for $-4<\alpha<0$, where we took $L=30$.
The derivative shows a singular increase at $\alpha=\alpha_c\approx -0.27$.
This value separates two phases, I for $\alpha_c<\alpha<0$ and II for $\alpha< \alpha_c$.
We can identify the phase I and II as the gSPT phase and a spin-2 Haldane phase, respectively.
In what follows, we amass evidences of this identification of quantum phases.

\subsection{Phase I: gSPT phase}

\begin{figure}[t!]
    \centering
    \includegraphics[width=\linewidth]{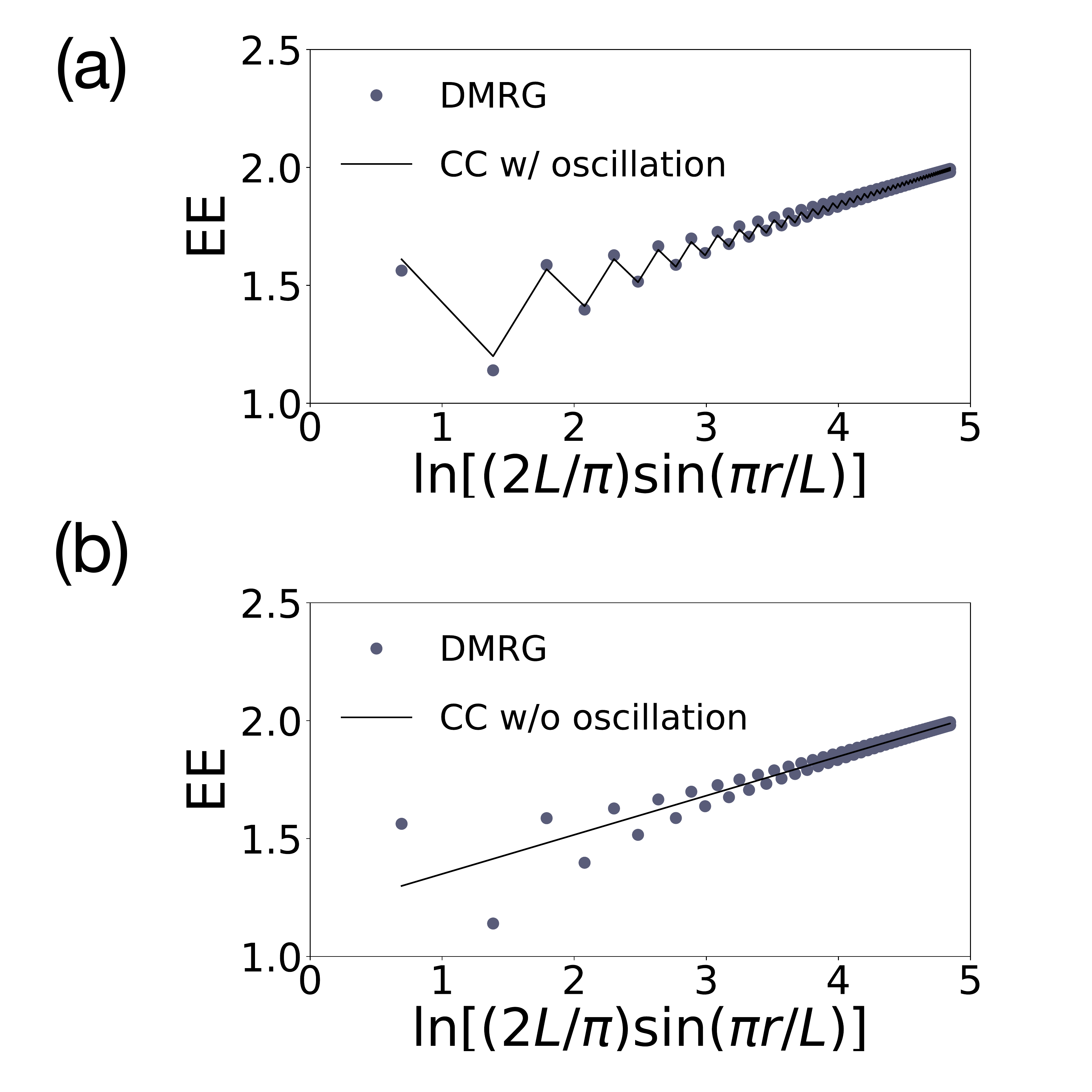}
    \caption{Entanglement entropy in gSPT phase.
    In both panels, gray balls are DMRG data with $L=200$, $J_t/J=0.2$, $J_\ell/J=0.1$, and $\alpha=-1.2$.
    Solid curves are fitting results of the DMRG data with the Calabrese-Cardy formula \eqref{CC} (a) with and (b) without the oscillating term \eqref{boundary_entropy} originating from boundary degrees of freedom.
    Both give the consistent fitting result of the central charge $c$ close to $c=1$.
    The former gives $c\approx 1.07$ and the latter gives $c\approx 0.995$. 
    }
    \label{fig:EE_log_gSPT}
\end{figure}

We numerically confirm that the phase I is the gSPT phase in two stages.
First, we show that the phase I is a critical phase with gapless excited states.
Second, we show that the ground state in the phase I has a nontrivial short-range entanglement.

\subsubsection{Entanglement entropy}

Let us show that the phase I is a critical phase described by a conformal field theory (CFT) with a central charge $c=1$.
Figure~\ref{fig:EE_log_gSPT} shows the entanglement entropy $S_E$ at $\alpha=-1.2$, where we take the system length $L=200$.
In the calculations of the entanglement, we fix the total magnetization to $\braket{S_{\text{tot}}^z}=1$ to minimize the boundary effects into the entanglement entropy.
We confirmed that the ground-state energies in $\braket{S_{\text{tot}}^z}=1,0,-1$ sectors are well degenerate.

If the conformal symmetry emerges at low energies, the entanglement entropy of the ground state with the OBC follows a so-called Calabresse-Cardy formula~\cite{CalabreseCardy2004},
\begin{align}
    S_E = \frac c6 \ln \biggl[\frac{2L}{\pi}\sin\biggl(\frac{\pi r}{L}\biggr)\biggr] + \ln g + a,
    \label{CC}
\end{align}
with a constant $a$.
The site $r=1,2,\cdots, L-1$ in Eq.~\eqref{CC} represents the bond between the $r$th and ($r+1$)th tetrahedra.
Note that we included a boundary entropy $\ln g$~\cite{Affleck1991_boundary_entropy,Laflorencie2006_EE} in Eq.~\eqref{CC}.
The boundary entropy shows the following site dependence
\begin{align}
    \ln g = b (-1)^x \biggl[\frac{L}{\pi}\sin\biggl(\frac{\pi r}{L}\biggr)\biggr]^{-1}.
    \label{boundary_entropy}
\end{align}
In the spin-1/2 XXZ chain, the constant $b$ takes a universal value $b=-1$~\cite{Laflorencie2006_EE}.
Considering the complexity of our system, we regard $b$ as a free parameter and determine $a$, $b$, and $c$ by comparing Eq.~\eqref{CC} with the DMRG data.
In Fig.~\ref{fig:EE_log_gSPT}~(a), we fit Eq.~\eqref{CC} with the DMRG data by changing $a$, $b$, and $c$.
We then obtain optimal values $(a,b,c)=(1.13124304,\, -0.35605475,\,  1.06698522)$.
The central charge $c\approx 1.07$ is close to the value $c=1$.
The Tomonaga-Luttinger liquid (TLL) is the most likely candidate of the $c=1$ CFT in quantum spin chains~\cite{giamarchi_book}.

We can take an alternative approach to estimate the central charge $c$.
That is, we ignore the data close to boundary and fit the DMRG data with the Calabrese-Cardy formula \eqref{CC} without the oscillating boundary term  (i.e., $b=0$).
In Fig.~\ref{fig:EE_log_gSPT}~(b), we discarded the first six data $r=1,2,\cdots,6$ and fit the remaining data with Eq.~\eqref{CC} with $b=0$.
Note that we plotted the DMRG data for $r=1,2,\cdots, L/2$ in Fig.~\ref{fig:EE_log_gSPT} because of the reflection symmetry $r\to L-r$ of the tetrahedral spin chain with the OBC.
We then obtain $(a,c)=(1.18421471,\, 0.99530899)$.
Again, we find the central charge $c\approx 0.995$ close to $c=1$.
Therefore, we conclude that the phase I is described by the $c=1$ CFT.

\begin{figure}[t!]
    \centering
    \includegraphics[width=\linewidth]{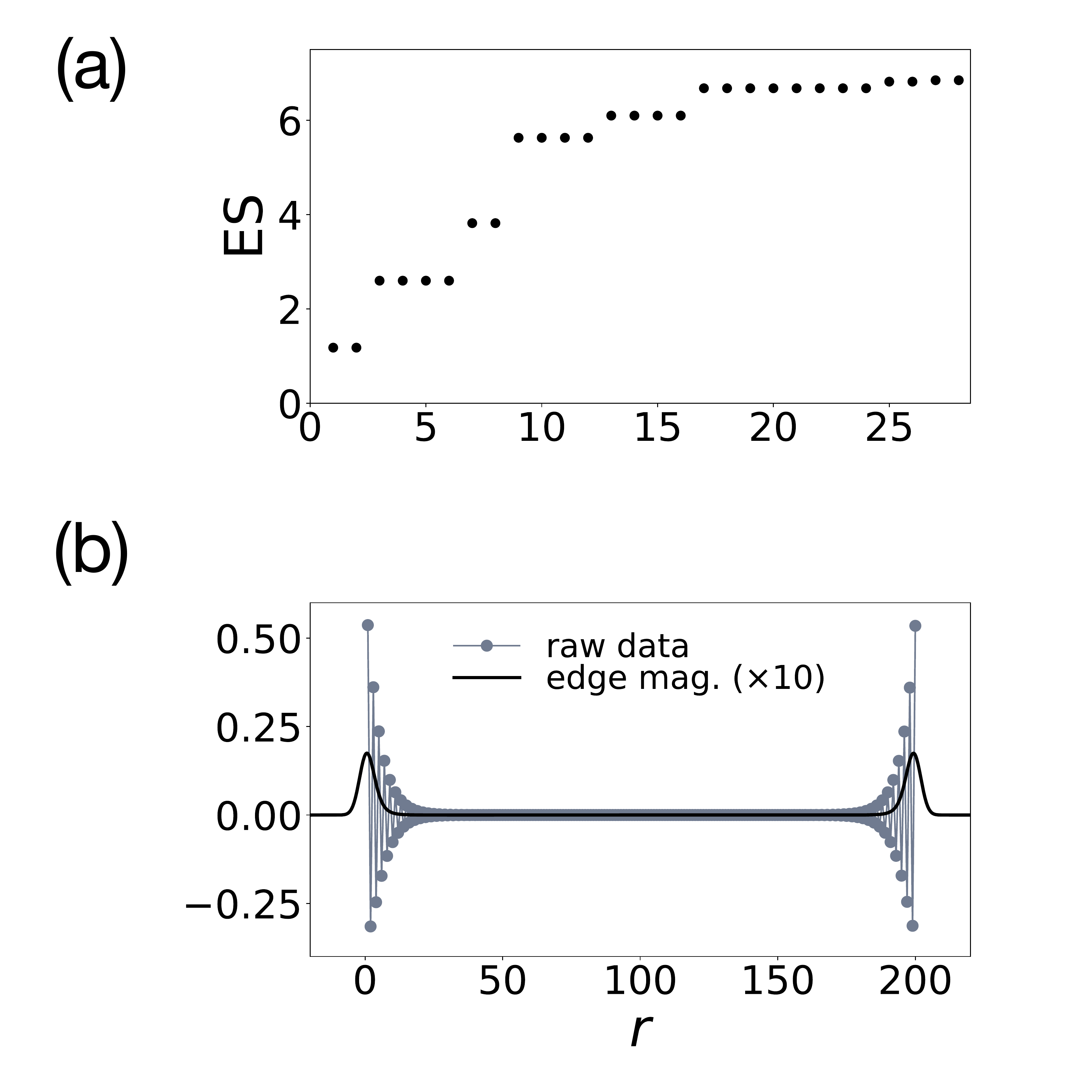}
    \caption{(a) Entanglement spectrum $\{\mu_i=-\ln(\lambda_i^2)\}_{i=1,2,\cdots}$ at central bond in gSPT phase ($\alpha=-1.2$ and $L=200$).
    $\lambda_i$ denotes the Schmidt eigenvalue.
    We label $\lambda_i$ with $i$ in ascending order of $\mu_i$.
    The horizontal axes show $i$ and $\lambda_i$, respectively.
    We can find clear even-fold degeneracy of the entanglement spectrum.
    (b) Magnetization density $S_{r,\text{tot}}^z$ (gray balls) and edge magnetization $m^z(r)$ (solid curve) plotted with respect to $r$.
    We derive the latter from the former through a Gaussian convolution~\cite{Watanabe2021_corner_mag}.
    Both quantities show the growth of the nonzero magnetization at both edges of the chain $r=1, L$.
    In particular, the latter shows a fractional quantization $M_{\text{left}}^z=M_{\text{right}}^z=1/2$, an evidence of the fractional $S=1/2$ edge spin of the AKLT state (see the main text.).
    }
    \label{fig:ES_gSPT}
\end{figure}

\subsubsection{Entanglement spectrum and edge magnetization}

The phase I has yet another interesting property about the entanglement.
Figure~\ref{fig:ES_gSPT}~(a) shows the entanglement spectrum $\{\mu_i\}_{i=1,2,\cdots}$ at $\alpha=-1.2$ with $L=200$ in the $\braket{S_{\text{tot}}^z}=1$ sector.
The eigenvalue $\mu_i=-\ln (\lambda_i^2)$ is obtained from the Schmidt eigenvalue $\lambda_i$.
We measured the entanglement at the center bond $r=L/2$ of the spin chain.
The entanglement spectrum exhibits a clear even-fold degeneracy.
The even-fold degeneracy implies that the ground state is a short-range-entangled VBS state similar to a spin-1 Haldane state.

The spin-1 Haldane state is the unique gapped ground state of the spin-1 Heisenberg antiferromagnetic chain with the PBC~\cite{Haldane1983a,Haldane1983b,Affleck1989_review_Haldane_gap} and one of the best-known gapped SPT states~\cite{Pollmann2010_Haldane_SPT,Pollmann2012_Haldnae_SPT,Tasaki2020_textbook}.
According to the bulk-edge correspondence in the gapped SPT phase, the nontrivial entanglement spectrum indicates the existence of nontrivial edge states.
Indeed, it is well known that the spin-1 Haldane state with the OBC hosts two symmetry-protected edge spins with the fractional spin quantum number $S=1/2$~\cite{AKLT1987}.

The ground state in our gSPT phase also shows edge states akin to the fractional $S=1/2$ edge state of the spin-1 Haldane state.
Gray balls in Fig.~\ref{fig:ES_gSPT}~(b) represent the spatial distribution $\braket{S_r^z}$ of the magnetization per tetrahedron:
\begin{align}
    \braket{S_r^z}=\sum_{n=1}^4 \braket{S_{r,n}^z}.
    \label{mag_dens}
\end{align}
The magnetization density \eqref{mag_dens} is nonzero around the edges of the chain while it is flatly zero deep inside the bulk.

We can show that the nonzero magnetization localized around the edges indeed represents the fractional $S=1/2$ edge state as follows.
Recently, a magnetic analog of the bulk electric polarization was proposed to characterize gapped quantum phases~\cite{Watanabe2021_corner_mag}.
This magnetic analog is called an edge (corner) magnetization in one-dimensional (two- or higher-dimensional, respectively) spin systems.
While Ref.~\cite{Watanabe2021_corner_mag}  focused on topologically trivial phases,
later, the edge magnetization turned out to also be relevant to SPT phases in spin chains.
In fact, one of the authors showed that the edge magnetizations $M_{\text{left}}^z$ and $M_{\text{right}}^z$ in the spin-1 Haldane state originate from the fractional $S=1/2$ edge spin~\cite{Furuya2021_edge_mag}.
Here, we denote the edge magnetizations on the left and right edges of the spin chain as $M_{\text{left}}^z$ and $M_{\text{right}}^z$, respectively.
The edge magnetizations in the spin-1 Haldane state are quantized as $M_{\text{left}}^z=\pm 1/2$ and $M_{\text{right}}^z=\pm 1/2$.
To fix the sign of the edge magnetization, we performed DMRG calculations in the $\braket{S_{\text{tot}}^z}=1$ sector.

We define the edge magnetizations $M_{\text{left}}^z$ and $M_{\text{right}}^z$ as follows.
First, we take a Gaussian convolution of the magnetization per tetrahedron $\braket{S_r^z}$:
\begin{align}
    m^z(r) = \sum_{r'=1}^L g(r-r') \braket{S_{r'}^z}, 
    \label{mz_convolution}
\end{align}
where $g(r)$ is a normalized Gaussian function parameterized with a positive constant $\lambda$:
\begin{align}
    g(r)= \frac{1}{\sqrt{2\pi \lambda^2}} \exp \biggl(-\frac{r^2}{2\lambda^2}\biggr).
\end{align}
The Gaussian convolution \eqref{mz_convolution} smooths the raw data of $\braket{S_r^z}$.
The solid curve of Fig.~\ref{fig:ES_gSPT}~(b) depicts $m^z(r)$ with $\lambda=5$.
The value of $\lambda>0$ can be arbitrary.

Next, based on the smoothed magnetization density \eqref{mz_convolution}, we define the edge magnetization $M_{\text{left}}^z$ ($M_{\text{right}}^z$) on the left (right) as an area swept by $m^z(r)$ when $r$ runs over the left (right, respectively) half of the system:
\begin{align}
    M_{\text{left}}^z = \int_{-\infty}^{(L+1)/2} dr \, m^z(r), \quad M_{\text{right}}^z = \int_{(L+1)/2}^\infty dr \, m^z(r).
    \label{edge_mag_def}
\end{align}
By construction, the sum of the edge magnetizations gives the total magnetization $M_{\text{left}}^z+M_{\text{right}}^z=
\braket{S_{\text{tot}}^z}$.

The DMRG raw data of Fig.~\ref{fig:ES_gSPT}~(b) gives
\begin{align}
    (M_{\text{left}}^z,\, M_{\text{right}}^z)=(0.4999999,\, 0.5000001).
    \label{edge_mag_num}
\end{align}
The edge magnetizations are well quantized to $M_{\text{left}}^z=M_{\text{right}}^z=1/2$.
We thus conclude that the phase I is the gSPT phase with the $c=1$ CFT criticality and the symmetry-protected $S=1/2$ edge spin.

\subsection{Phase II: spin-2 Haldane phase}

\begin{figure}[t!]
    \centering
    \includegraphics[width=\linewidth]{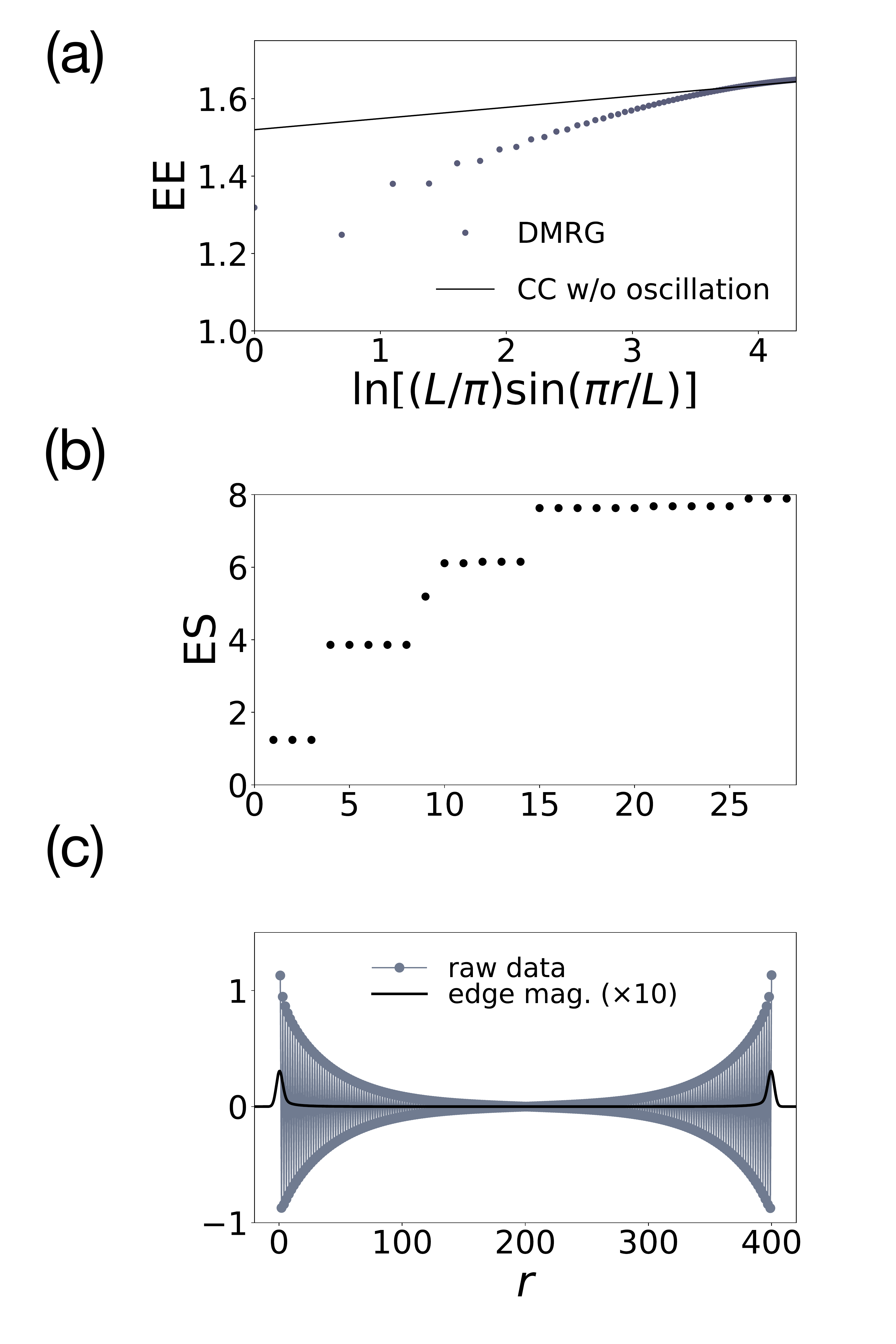}
    \caption{(a) Entanglement entropy in phase II. (b) Entanglement spectrum in phase II. (c) Edge magnetizations in phase II.
    In the three panels, filled balls are DMRG data with $L=400$, $J_t/J=0.2$, $J_\ell=0.1$, and $\alpha=-3.5$.
    The DMRG calculations are done with $\braket{S_{\text{tot}}^z}=2$ to minimize the boundary effects.
    The solid curve is the Calabrese-Cardy formula \eqref{CC} without the boundary term ($b=0$).
    The entanglement entropy does not fit well with the Calabrese-Cardy formula, implying that the phase II is gapped.
    The entanglement spectrum shows no special degeneracy in contrast to the phase I, implying that the phase II is topologically trivial.
    The edge magnetization implies that the phase II is the spin-2 Haldane one because of the quantized edge magnetizations, $(M_{\text{left}}^z,M_{\text{right}}^z)=(0.99999, 1.00001)$.
    }
    \label{fig:EE_log_ES_mag_H2}
\end{figure}

We now give our attention to the phase II.
Figure~\ref{fig:EE_log_ES_mag_H2} shows (a) the entanglement entropy, (b) the entanglement spectrum, and (c) the edge magnetization at $\alpha=-3.5$ with $L=350$ in the $\braket{S_{\text{tot}}^z}=2$ sector.
We obtained the solid curve of Fig.~\ref{fig:EE_log_ES_mag_H2}~(a) by  fitting the numerically obrtained entanglement entropy in a range $30<r<L$ with the Calabresse-Cardy formula \eqref{CC} with $b=0$.
We then obtained the result $(a,c)=(1.51995057,\, 0.17321388)$.
However, the Calabresse-Cardy formula \eqref{CC} does not fill well with the numerically obtained entanglement entropy in the range $30<r<L$.
Hence, the phase II is not critical and highly likely to be weakly gapped.
The entanglement spectrum of Fig.~\ref{fig:EE_log_ES_mag_H2}~(b) indicates that the phase II is topologically trivial.
Finally, Fig.~\ref{fig:EE_log_ES_mag_H2}~(c) shows the quantized edge magnetizations $(M_{\text{left}}^z,M_{\text{right}}^z)=(0.99999, 1.00001)$.
The DMRG calculations in Fig.~\ref{fig:EE_log_ES_mag_H2} are done under a constraint $\braket{S_{\text{tot}}^z}=2$ to make $M_{\text{left}}^z$ and $M_{\text{right}}^z$ positive.
We confirmed that the ground-state energies in the $\braket{S_{\text{tot}}^z}=2$, $1$, $0$, $-1$, and $-2$ sectors are degenerate.
Note that even though the edge magnetization stands, the edge state in the spin-2 Haldane phase is not protected by any symmetry as the nondegenerate entanglement spectrum indicates~\cite{Pollmann2010_Haldane_SPT,Pollmann2012_Haldnae_SPT,Tonegawa2011_spin2}.
With these discussions, we identify the phase II as the spin-2 Haldane phase.

\section{Perturbation theory}\label{sec:perturbation}

In this section, we identify the $c=1$ CFT in the phase I.
For this purpose, we split the Hamiltonian \eqref{H_tetra} into two parts:
\begin{align}
    \mathcal H
    &= \mathcal H_0 + V,
\end{align}
where $\mathcal H_r$ and $V$ are the intra-tetrahedron and inter-tetrahedron interactions, respectively:
\begin{subequations}
\begin{align}
    \mathcal H_0
    &= \sum_r h_r, 
    \label{H0_def} \\
    h_r 
    &= J(\bm{S}_{r,1}\cdot\bm{S}_{r,2} + \bm S_{r,2}\cdot \bm S_{r,3} + \bm S_{r,3}\cdot \bm S_{r,1})
    \notag \\
    &\qquad +\alpha J \bm S_{r,4}\cdot \bm T_r,
    \label{hr_def}
\end{align}
\end{subequations}

\noindent
and 
\begin{align}
    V 
    &= J_t \sum_{r=1}^L \bm S_{r,4}\cdot \bm T_{r+1} +J_\ell \sum_{r=1}^L \sum_{n=1}^3 \bm S_{r,n} \cdot \bm S_{r+1,n}.
    \label{V_def}
\end{align}
We regard $V$ as a perturbation to $\mathcal H_0$.
In other words, we regard $J_t/J$ and $J_\ell/J$ small perturbative parameters.

\subsection{Unperturbed ground states}

The unperturbed Hamiltonian \eqref{H0_def} describes mutually isolated tetrahedra.
Each tetrahedron is governed by the simple Hamiltonian \eqref{hr_def} paraphrased as
\begin{align}
    h_r = \frac{J(1-\alpha)}{2}\bm{T}_r^2
    +\frac{\alpha J}2\bm{S}_{r,\text{tot}}^2+\mathrm{const.}
    \label{hr_simplified}
\end{align}
This Hamiltonian consists only of $\bm{T}_r$ of the base triangle and the total spin $\bm{S}_{r,\text{tot}}=\bm{T}_r+\bm{S}_{r,4}=\sum_{n=1}^4\bm{S}_{r,n}$ of the tetrahedron.
The Hamiltonian \eqref{hr_simplified} conserves the spin quantum numbers
$T_r=1/2$ or $3/2$ of $\bm{T}_r$ and $S_{r,\text{tot}}=0,1$, or $2$ of $\bm{S}_{r,\text{tot}}$,  and the $z$ component $S_{r,\text{tot}}^z$ of $\bm{S}_{r,\text{tot}}$.
Every eigenstate is thus labeled by at least three parameters, which we denote $\ket{T_r, S_{r,\text{tot}},S_{r,\text{tot}}^z}$.

We can easily find a ground state of the Hamiltonian \eqref{hr_simplified} by solving the eigenvalue equation.
For $\alpha < -3$, the ground state of $h_r$ has $T_r=3/2$ and $S_{r,\text{tot}}=2$, where the ground state is five-fold degenerate: $\ket{\frac 32, 2, m}$ with $m=2,1,0,-1,-2$.
For $-3<\alpha<0$, the ground state has $T_r=1/2$ and $S_{\text{tot}}=1$, but is six-fold degenerate: $\ket{\frac 12, 1, m, \chi}$ with $m=1,0,-1$ and $\chi=\pm 1$.
Here, the ground state admits the additional parameter $\chi=\pm 1$ that represents the eigenvalue of a scalar chirality,
\begin{align}
    \chi_r &=\frac{4}{3\sqrt{3}}(\bm{S}_{r,1}\cdot\bm{S}_{r,2}\times\bm{S}_{r,3} + \bm{S}_{r,2}\cdot\bm{S}_{r,3}\times\bm{S}_{r,1} 
    \notag \\
    &\qquad 
    + \bm{S}_{r,3}\cdot\bm{S}_{r,1}\times\bm{S}_{r,2}).
    \label{chirality_def}
\end{align}
This scalar chirality is defined on the base triangle of the tetrahedron but can also be identified as the chirality of the tetrahedron since $\alpha\not=1$ [Fig.~\ref{fig:tetrahedron}~(b)].
The chirality \eqref{chirality_def} is a central figure in this paper.
We determined the factor $4/3\sqrt{3}$ from explicit representations of the eigenstates.

Generally, chirality degrees of freedom affect magnetism when the magnetic configuration is canted and noncollinear~\cite{Kawamura1998_review}.
The geometrical frustration can trigger such magnetic configurations with nonzero scalar chirality.
In our tetrahedron, the chirality must be incorporated for $-3<\alpha<0$, where the geometrically frustrated interaction $J(\bm S_{r,1}\cdot \bm S_{r,2} + \bm S_{r,2}\cdot \bm S_{r,3} + \bm S_{r,3}\cdot \bm S_{r,1})$ in the base triangle leads to a canted state with $T_r^z=\pm 1/2$.
Let us denote the eigenstate of the base triangle as $\ket{T_r^z, \chi}_{123}$. 
For $T_r^z=1/2$, they are given by
\begin{align}
    \ket{\tfrac 12,+}_{123}
    &= \frac{1}{\sqrt{3}} (
    \ket{\downarrow_1\uparrow_2\uparrow_3}+\omega \ket{\uparrow_1\downarrow_2\uparrow_3} + \omega^{-1}\ket{\uparrow_1\uparrow_2\downarrow_3}), \\
    \ket{\tfrac{1}{2},-}_{123}
    &= \frac{1}{\sqrt{3}} (
    \ket{\downarrow_1\uparrow_2\uparrow_3}+\omega^{-1} \ket{\uparrow_1\downarrow_2\uparrow_3} + \omega \ket{\uparrow_1\uparrow_2\downarrow_3}),
\end{align}
with a complex constant $\omega=\exp(i\frac{2\pi}{3})$ and an eigenstate $\ket{s_1s_2s_3}$ with $S_{r,n}^z=s_n$ for $n=1,2,3$.
It is easy to verify that these $\ket{\tfrac 12, \chi}_{123}$ satisfy
\begin{align}
    \chi_r \ket{\tfrac 12, \chi}_{123} = \chi \ket{\tfrac 12, \chi}_{123}.
\end{align}
We can build the eigenstate $\ket{\frac 12, 1, m, \chi}$ by combining $\ket{\pm\frac 12, \chi}_{123}$ and eigenstates $\ket{\uparrow}_4$ and $\ket{\downarrow}_4$ of $S_{r,4}^z$ ferromagnetically [Eq.~\eqref{GS_tetrahedron}].

The ground state of $h_r$ switches at $\alpha=-3$, a close value to the transition point $\alpha_c\approx -2.7$ obtained from Fig.~\ref{fig:tetrahedron}~(c).
We may regard $\alpha_c=-3$ as the zeroth-order perturbation value of the phase transition point with respect to $J_t/J$ and $J_\ell/J$.

\subsection{Degenerate perturbation theory in phase I}

We focus ourselves on the phase I in the range $\alpha_c<\alpha<0$ and
take the $J_t$ and $J_\ell$ interactions into account.
These interactions lift the degeneracy of the unperturbed ground state.
To describe the perturbation theory, we prepare an effective $S=1$ spin $\bm{s}_r$ and an effective $S=1/2$ pseudospin $\bm{t}_r$ from the sixfold degenerate ground states $\ket{\frac{1}{2},1,m,\chi}$ of $h_r$.
The $z$ components of these spins are defined as
\begin{align}
    s_r^z\ket{\tfrac{1}{2},1,m,\chi} 
    &=m\ket{\tfrac{1}{2},1,m,\chi}, \\ t_r^z\ket{\tfrac{1}{2},1,m,\chi}
    &=\frac{\chi}{2}\ket{\tfrac{1}{2},1,m,\chi}.
\end{align}
The raising operators $s_r^+$ and $t_r^+$ increase the $z$ components $m$ and $\chi/2$ by $1$, respectively.
The lowering operators are defined similarly.
We can derive a low-energy effective representation of the spin operator $\bm S_{r,n}$ by projecting the operator into the low-energy Hilbert subspace (see Appendix.~\ref{app:pseudospin}):
\begin{align}
    \bm S_{r,n} &\sim \frac 16 (1-2\omega^{n-1}t_r^+-2\omega^{-(n-1)}t_r^-)\otimes \bm s_r, 
    \label{Sn_low-energy} \\
    \bm S_{r,4} &\sim \frac 12 \bm s_r,
    \label{S4_low-energy}
\end{align}
for $n=1,2,3$.
Here, the symbol $\sim$ means an approximate identity in the low-energy Hilbert subspace well below the spin gap.

The low-energy physics of the tetrahedral spin chain \eqref{H_tetra} is effectively described by $\bm{s}_r$ and $\bm{t}_r$.
The perturbation expansion with respect to $J_t$ and $J_\ell$ leads to the low-energy effective Hamiltonian:
\begin{align}
    \mathcal{H}_{\text{eff}}
    &\approx \frac{3J_t+J_\ell}{12} \sum_{r=1}^L\bm{s}_r\cdot \bm{s}_{r+1}
    \notag \\
    &\quad
    +\frac{J_\ell}{12} \sum_{r=1}^L (t_r^xt_{r+1}^x+t_r^yt_{r+1}^y)\bm{s}_r\cdot\bm{s}_{r+1},
    \label{H_eff}
\end{align}
where the second- and higher-order terms are discarded.
Within the first-order approximation, we obtain the effective Hamiltonian \eqref{H_eff} by replacing the four spins $\bm S_{r,n}$ ($n=1,2,3,4$) with their effective low-energy representations, Eqs.~\eqref{Sn_low-energy} and \eqref{S4_low-energy}.

We can confirm that the right hand side of the Hamiltonian \eqref{H_eff} indeed has the gSPT ground state.
The first line of Eq.~\eqref{H_eff} is the spin-1 Heisenberg antiferromagnetic interaction that trivially opens the spin gap.
When $J_t$ is larger than $J_\ell$, the second line of Eq.~\eqref{H_eff} does not close the spin gap.
Note that the trivial spin-gap opening keeps the lattice translation symmetry.
The lattice translation symmetry makes the ground-state expectation value $\braket{\bm{s}_r\cdot\bm{s}_{r+1}}=C_1$ spatially uniform.
In other words, the constant $C_1$ is independent of the unit-cell index $r$.
Then, we may further approximate Eq.~\eqref{H_eff} as
\begin{align}
    \mathcal{H}_{\text{eff}}
    &\approx \frac{3J_t+J_\ell}{12} \sum_{r=1}^L\bm{s}_r\cdot \bm{s}_{r+1} 
    \notag \\
    &\qquad +\frac{C_1J_\ell}{12} \sum_{r=1}^L (t_r^xt_{r+1}^x+t_r^yt_{r+1}^y),
    \label{H_eff_decoupled}
\end{align}
where the spin-1 $\bm{s}_r$ and the spin-1/2 $\bm{t}_r$ are effectively decoupled.
The approximation \eqref{H_eff_decoupled} holds when we are concerned with the ground state and low-energy excitations well below the spin gap.
To know the precise value of $C_1$, we need a self-consistent determination process to include the effects of the pseudospin $\bm t_r$.
However, the precise value is not important.
Irrespective of sign of $C_1J_\ell$, the second line of \eqref{H_eff_decoupled} represents that the pseudospin $\bm t_r$ forms the TLL ground state.
This TLL is  known as the chirality liquid~\cite{Okunishi2012_spintube_chiral}.

Since the $S=1/2$ pseudospin $\bm t_r$ follows the XY chain in Eq.~\eqref{H_eff_decoupled}, we can bosonize the pseudospin as
\begin{align}
    t_r^z &= \frac{1}{\pi} \partial_x \phi + a_1 \sin(2\phi),
    \label{tz_bosonized} 
    \\
    t_r^+ &= e^{i\theta} [(-1)^r b_0 + b_1 \sin (2\phi)],
    \label{t+_bosonized}
\end{align}
where $a_1$, $b_0$, and $b_1$ are constants, $\phi(r)$ is a U(1) compact boson field, and $\theta(r)$ is its cannonical conjugate that satisfy a commutation relation $[\phi(x), \theta(y)]=\pi i\Theta(x-y)$.
Here, $\Theta(z)$ is the Heaviside step function with $\Theta(0)=1/2$.
We can bosonize the XY intearction of the pseudospin similarly to that of the authentic spin~\cite{giamarchi_book}:
\begin{align}
    \mathcal H_{\text{eff}}
    &\approx \frac v{2\pi} \int dr \{(\partial_r\theta)^2+ (\partial_r\phi)^2 \},
    \label{H_eff_bosonized}
\end{align}
where we dropped the $\bm s_r$ part because it only generates high-energy excitations.
The Hamiltonian \eqref{H_eff_bosonized} describes the $c=1$ CFT of the free boson $\phi$~\cite{cft_yellowbook}.
We thus identify the ground state of the phase I as the chirality liquid with symmetry-protected edge spins.

However, our argument thus far is only half the battle.
The effective Hamiltonian \eqref{H_eff} and its final form \eqref{H_eff_bosonized} hold only within the first-order approximation of $J_t/J$ and $J_\ell/J$.
Higher-order terms of the perturbation expansion yield various interactions.
To conclude that the phase I is the gSPT phase, we need to confirm that the higher-order corrections to the effective Hamiltonian \eqref{H_eff} do not trivially open the gap.
For example, if the effective Hamiltonian should admit an interaction $\sum_r t_r^x \propto \int dr \, \cos\theta$, the chirality liquid would immediately acquire the excitation gap and make the ground state unique and gapped.
Fortunately, we have a complete list of the relevant interactions in the sense of the renormalization group: $\cos(p\theta)$, $\sin(p\theta)$ for $p=1,2$, $\cos(2\phi)$,  $\sin(2\phi)$, $\partial_x\phi$, and $\partial_x\theta$.
Carefully analyzing what symmetry forbids these relevant interactions, we can discuss the ingappability of the chirality liquid.
Reference~\cite{Okunishi2012_spintube_chiral} adopts this strategy to discuss the ingappability of the chirality liquid of an $S=1/2$ three-leg spin tube on a $1/3$ magnetization plateau.
In the next section, we adopt a more generic approach to show the ingappability and later come back to the specific case of the chirality liquid.

\section{LSM theorem for discrete symmetries}\label{sec:LSM}

In this section, we show the ingappability of the phase I under symmetries from the viewpoint of the Lieb-Schultz-Mattis (LSM) theorem for discrete symmetries.
The LSM theorem is a well-known no-go theorem about the ingappability of quantum phases~\cite{lsm,affleck1988_lsm,Oshikawa2000_LSM,hastings2004_lsm}.
As a fundamental principle that determines the fate of the quantum phase, the LSM theorem has long been discussed and applied to a wide variety of many-body systems~\cite{Affleck1986_LSM,affleck1988_lsm,hastings2004_lsm,Hastings2005_LSM,Oshikawa1997_plateau_OYA,Oshikawa2000_LSM,Oshikawa2003_LSM_topo,Cho2017_LSM,Watanabe2015_LSM_PNAS,Cheng2019_LSM,Furuya2017_spc,Yao2019_spc,Ogata2019_LSM_discrete}.
The LSM theorem and ingappability are essential to characterize the gSPT phase~\cite{hidaka2022_gaplessSPT}.

We give our attention to a $\mathbb Z_3$ symmetry, a $\mathbb Z_2^T$ symmetry, and the lattice translation symmetry ($\bm S_{r,n}\to \bm S_{r+1,n}$).
Here, $\mathbb{Z}_3$ represents the $\mathbb{Z}_3$ rotation symmetry around an axis piercing spins $\bm S_{r,4}$ for $r=1,2,\cdots, L$ [the dashed blue line of Fig.~\ref{fig:tetrahedron}~(b)].
\begin{align}
    \mathcal{R}\bm{S}_{r,n}\mathcal{R}^{-1}=\bm{S}_{r,n'},
    \label{Z3_rot_def}
\end{align}
with 
\begin{align}
    n' &= \left\{
    \begin{array}{ccc}
    2, && (n=1), \\
    3, && (n=2), \\
    1, && (n=3), \\
    4, && (n=4).
    \end{array}
    \right.
    \label{n'}
\end{align}
$\mathbb{Z}_2^T$ represents the time-reversal symmetry, 
\begin{subequations}
\begin{align}
    \mathcal{T}\bm{S}_{r,n}\mathcal{T}^{-1}
    & =-\bm{S}_{r,n}, \\
    \mathcal{T}i\mathcal{T}^{-1}
    &=-i.
\end{align}
\label{TR_def}
\end{subequations}

\noindent
Our argument is basically a derivative of a method used in Ref.~\cite{yao2021_LSM_STBC} to show a LSM theorem for discrete symmetries.

\subsection{Gauging global $\mathbb Z_3$ symmetry}

The LSM theorem is closely related to an 't Hooft anomaly~\cite{Hsieh2014_SPT_orientifold,Furuya2017_spc}.
The 't Hooft anomaly is an inconsistency between plural global symmetries that appears when gauging one of the global symmetries, i.e., when promoting the symmetry to a local gauge symmetry~\cite{thooft1980_anomaly,Kapustin2014_anomaly}.
In the presence of the local gauge symmetry, the excitation spectrum is invariant under a local gauge transformation.
The local gauge transformation is also referred to as a local gauge twist~\cite{Hatsugai2006_local_BP,Chepiga2013_local_BP,Chen2015_gauging_TR,Kariyado2018_ZN_BP,Maruyama2018_ZN_BP}.
The local gauge symmetry motivates us to consider a closed boundary condition accompanied by a local gauge twist on its seam instead of the PBC.
Let us call this boundary condition a symmetry-twisted boundary condition (STBC) following Ref.~\cite{yao2021_LSM_STBC}.
If we gauge the global $\mathbb{Z}_3$ symmetry, the corresponding STBC, which we call a $\mathbb{Z}_3$ STBC, is defined as 
\begin{align}
    \bm S_{r+L,n} &= \bm S_{r,n'},
\end{align}
with $n'$ of Eq.~\eqref{n'}.
Only when we pass over the seam of the system, we locally feel the $\mathbb Z_3$ spatial rotation.

The Hamiltonian with the $\mathbb Z_3$ STBC is 
\begin{align}
    \mathcal{H}_R
    &=J\sum_{r=1}^L  (\bm S_{r,1}\cdot \bm S_{r,2} + \bm S_{r,2}\cdot \bm S_{r,3} +\bm S_{r,3}\cdot \bm S_{r,1})
    \notag \\
    &\qquad +\alpha J\sum_{r=1}^L \bm{S}_{r,4}\cdot\bm{T}_{r}
    \notag \\
    &\qquad +J_t\sum_{r=1}^{L}\bm{S}_{r,4}\cdot\bm{T}_{r+1} + J_\ell\sum_{r=1}^{L-1}\sum_{n=1}^3\bm{S}_{r,n}\cdot\bm{S}_{r+1,n}
    \notag \\
    &\qquad +J_\ell (\bm S_{L,1}\cdot \bm S_{1,2} + \bm S_{L,2} \cdot \bm S_{1,3} + \bm S_{L,3}\cdot \bm S_{1,1}).
    \label{H_STBC}
\end{align}
The last interaction on the seam of the $\mathbb Z_3$ STBC gets twisted (rotated) by the local $\mathbb{Z}_3$ gauge transformation.
With the $\mathbb Z_3$ STBC, the lattice translation $\tilde{T}_1$ is given by
\begin{align}
    \tilde{T}_1 \bm{S}_{j,n}\tilde{T}_1^{-1}
    &= \left\{
    \begin{array}{ccc}
    \bm{S}_{j+1,n}, && (j\not=L), \\
    \bm{S}_{1,n'}, && (j=L),
    \end{array}
    \right.
\end{align}
with $n'$ of Eq.~\eqref{n'}.
This lattice translation operator admits a simple representation,
\begin{align}
    \tilde{T}_1=R_1T_1=T_1R_L.
    \label{T1_STBC_def}
\end{align}
Here, $T_1$ is the lattice translation in the PBC and $R_j$ is the $\mathbb{Z}_3$ rotation that locally acts on the $j$th tetrahedron, that is,
\begin{align}
    T_1\bm S_{j,n} T_1^{-1} &= \bm S_{j+1,n},
    \label{T1_PBC_def} \\
    R_j \bm S_{k,n} R_j^{-1}
    &= \delta_{j,k}\bm{S}_{j,n'}+(1-\delta_{j,k})\bm{S}_{j,n}.
    \label{Rj_def}
\end{align}
The global $\mathbb{Z}_3$ symmetry is generated by $\mathcal{R}=R_1R_2\cdots R_L$.
The lattice translation \eqref{T1_STBC_def} is a straightforward generalization of that of Ref.~\cite{yao2021_LSM_STBC}
We can confirm that $[\tilde{T}_1,\mathcal{H}_R]=0$ and $[T_1,\mathcal{H}_R]\not=0$.
In addition, the Hamiltonian \eqref{H_STBC} has the time-reversal symmetry, $[\mathcal{T},\mathcal{H}_R]=0$.
It is noteworthy that the $\mathbb Z_3$ rotation and the time-reversal symmetries are commutative,
\begin{align}
    [\mathcal T, \mathcal R] = 0.
\end{align}

\subsection{Ingappability}

With these preparations, we show by contradiction the ingappability of the phase I under the $\mathbb Z_3\times \mathbb Z_2^T$ and the lattice translation symmetries.
A key observation is that the unique gapped ground state is insensitive to the local gauge twist.
That is, when the ground state of the Hamiltonian \eqref{H_tetra} with the PBC is unique and gapped, so is the ground state of the Hamiltonian \eqref{H_STBC} with the STBC.
This insensitivity of the ground state to the local gauge twist is not rigorously proven but a physically sound conjecture~\cite{furuya2019_LSM_checkerboard,yao2021_LSM_STBC}.
Hence, to show the anomaly between the $\mathbb{Z}_3\times\mathbb{Z}_2^T$ symmetry and the translation symmetry, it suffices to find a contradiction by assuming the unique gapped ground state under the STBC.

Let $\ket{\psi_0}_R$ be the unique gapped ground state of the Hamiltonian with the STBC \eqref{H_STBC}.
We represent the ground state as
\begin{align}
    \ket{\psi_0}_R=\sum_{i_1,i_2,\cdots,i_L}c_{i_1i_2\cdots i_L}\ket{i_1i_2\cdots i_L},
    \label{psi0_R}
\end{align}
with $c_{i_1i_2\cdots i_L}\in\mathbb{C}$ and a product state $\ket{i_1i_2\cdots i_L}=\ket{i_1}\otimes\ket{i_2}\otimes\cdots\otimes\ket{i_L}$ with $i_r=1,2,\cdots,6$ for all $r=1,2,\cdots,L$.
We use a basis,
\begin{subequations}
\begin{align}
    \ket{\tfrac{1}{2},1,1,+} &=\ket{1}, \\
    \ket{\tfrac{1}{2},1,0,+} &=\ket{2}, \\
    \ket{\tfrac{1}{2},1,-1,+} &=\ket{3}, \\
    \ket{\tfrac{1}{2},1,1,-} &=\ket{4}, \\
    \ket{\tfrac{1}{2},1,0,-} &=\ket{5}, \\
    \ket{\tfrac{1}{2},1,-1,-} &=\ket{6}.
\end{align}
\label{basis_def}
\end{subequations}

\noindent 
We can take the ground state \eqref{psi0_R} as an eigenstate of $\tilde{T}_1$: 
\begin{align}
    \tilde{T}_1\ket{\psi_0}_R=e^{iP_0}\ket{\psi_0}
\end{align}
with $P_0\in[0,2\pi)$.

Now we consider a state 
\begin{align}
    \ket{\psi'_0}_R=\mathcal{T}\ket{\psi_0}_R.
\end{align}
If $\ket{\psi_0}_R$ was the unique gapped ground state, $\ket{\psi'_0}_R$ would be identical to $\ket{\psi_0}_R$ except for a U(1) factor since $[\mathcal H_R,\mathcal T]=0$.
In what follows, we show that $\ket{\psi'_0}_R$ cannot be identical to $\ket{\psi_0}_R$ by looking into a $\tilde{T}_1$ eigenvalue of $\ket{\psi'_0}_R$:
\begin{align}
    \tilde T_1\ket{\psi'_0}_R = e^{iP'_0}\ket{\psi'_0}_R.
    \label{T1_psi'_0}
\end{align}
It suffices to show $P'_0\not= P_0 \mod 2\pi$.
To calculate the left hand side of Eq.~\eqref{T1_psi'_0}, we need to know how the time-reversal $\mathcal{T}$ and the local $\mathbb{Z}_3$ spatial rotation $R_r$ act on $\ket{i_r}$.
They act on $\ket{i_r}$ as [Eqs.~\eqref{TR_proj} and \eqref{Z3_proj}]
\begin{align}
    \mathcal{T}\ket{i_r} &=\sum_{j_r=1}^6 \Bigl(-2t_r^x\otimes \exp(i\pi s_r^z)\Bigr)_{i_rj_r}\ket{j_r},
    \label{TR_eff}
    \\
    R_r\ket{i_r} &= \sum_{j_r=1}^6 \biggl(\exp\biggl(i\frac{4\pi}{3}t_r^z\biggr)\otimes\mathbbm{1}_3\biggr)_{i_rj_r}\ket{j_r}.
    \label{rot_eff}
\end{align}
The time reversal $\mathcal T$ turns into the $\pi$ rotations:
\begin{subequations}
\begin{align}
    \mathcal T (s_r^x, s_r^y, s_r^z) \mathcal T^{-1}
    &\sim (s_r^x, -s_r^y, -s_r^z), \\
    \mathcal T (t_r^x, t_r^y, t_r^z) \mathcal T^{-1}
    &\sim (t_r^x, -t_r^y, -t_r^z).
\end{align}
\label{TR_pi_rot}
\end{subequations}
\noindent
The global $\mathbb Z_3$ spatial rotation $\mathcal R=R_1R_2\cdots R_L$ acts only on the pseudospin:
\begin{subequations}
\begin{align}
    \mathcal R t_r^z \mathcal R^{-1} &\sim t_r^z, \\
    \mathcal R t_r^+ \mathcal R^{-1} &\sim \omega t_r^+.
\end{align}
\label{Z3_rot_eff}
\end{subequations}

The symmetry operations \eqref{TR_eff} and \eqref{rot_eff} lead to
\begin{align}
    R_r\mathcal T\ket{i_r} = \mathcal TR_r^{-1}\ket{i_r}.
    \label{RT_TR-1}
\end{align}
The relation \eqref{RT_TR-1} comes from the anticommutation relation
\begin{align}
    t_j^x t_j^z = -t_j^z t_j^x,
\end{align}
of the spin-1/2 operator $\bm t_j$.
The left hand side of Eq.~\eqref{T1_psi'_0} thus becomes
\begin{align}
    \tilde{T}_1\ket{\psi'_0}_R
    &=e^{iP_0}R_1^{-1}\ket{\psi'_0}_R.
    \label{T1_eigen_eq}
\end{align}
Equations~\eqref{T1_psi'_0} and \eqref{T1_eigen_eq} lead to
\begin{align}
    R_1\ket{\psi'_0}_R &= e^{i(P_0-P'_0)}\ket{\psi'_0}_R.
\end{align}
That is, $\ket{\psi'_0}_R$ is an eigenstate of the local $\mathbb Z_3$ rotation operator $R_1$.
Since $R_1$ has the eigenvalue $\exp(\pm \frac{2\pi i}{3})$ [Eq.~\eqref{rot_eff}], we obtain
\begin{align}
    P'_0 = P_0 \pm \frac{2\pi}{3} \mod 2\pi.
\end{align}
Hence, $\ket{\psi'_0}_R$ and $\ket{\psi_0}_R$ are the doubly degenerate ground states that spontaneously breaks the time-reversal symmetry, which contradicts the assumption of the unique gapped ground state.
Therefore, we conclude the anomaly between $\mathbb{Z}_3\times\mathbb{Z}_2^T$ symmetry and the lattice translation symmetry.
In other words, the tetrahedral chain \eqref{H_tetra} cannot have the unique gapped ground state in the presence of the $\mathbb{Z}_3\times\mathbb{Z}_2^T$ and the lattice translation symmetries as long as the chirality degrees of freedom govern the low-energy physics.

Our argument about the anomaly relies only on symmetries of the $S=1$ spin $\bm s_r$ and the $S=1/2$ pseudospin $\bm t_r$.
However, there is a precondition that the low-energy Hilbert subspace is locally spanned by the six states \eqref{basis_def}, where we implicitly assume the U(1) spin-rotation symmetry around $S^z$.
Should the U(1) spin-rotation symmetry be absent, those low-energy states would be mixed with other local eigenstates with $S_{r,\text{tot}}\not=1$ and would violate the above precondition.
Therefore, we need the $\mathbb Z_3\times \mathbb Z_2^T$, lattice translation, and U(1) spin-rotation symmetries to make the gSPT phase ingappable.

\subsection{Discussions}

\subsubsection{Applicability}

Our argument in the previous subsections generically holds when the unit cell contains the six local low-energy state $\ket{i_r}$ \eqref{basis_def} irrespective of how neighboring tetrahedra are coupled.
Hence, the ingappability holds even when we replace the interaction \eqref{V_def} with a completely different interaction unless it violates one of the $\mathbb Z_3$, the $\mathbb Z_2^T$, the lattice translation, and the U(1) spin-rotation symmetries.

\subsubsection{Specific case: chirality liquid}

However, since we are concerned with the chirality liquid, we should go back to this specific case and translate the generic result into the language of the chirality liquid \eqref{H_eff_bosonized}.

Let us check that the relevant interactions, $\cos(2\phi)$, $\sin(2\phi)$, $\cos(p\theta)$, $\sin(p\theta)$ for $p=1,2$, $\partial_x\phi$, and $\partial_x\theta$,
of the $c=1$ CFT \eqref{H_eff_bosonized} are all forbidden by the above-mentioned symmetries.
The action of the $\mathbb Z_3$ symmetry is simple.
The global $\mathbb Z_3$ rotation affects only the pseudospin [Eq.~\eqref{Z3_rot_eff}].
In terms of the boson fields \eqref{tz_bosonized} and \eqref{t+_bosonized}, they read as
\begin{align}
    \mathcal R \phi(r) \mathcal R^{-1} &= \phi(r), \\
    \mathcal R \theta(r) \mathcal R^{-1} &= \theta(r) + \frac{2\pi}{3}.
\end{align}
The $\mathbb Z_3$ symmetry thus forbids $\cos(p\theta)$ and $\sin(p\theta)$ for $p=1,2$.

The lattice translation acts on
\begin{align}
    T_1 \bm s_{r} T_1^{-1} &= \bm s_{r+1}, \\
    T_1 \bm t_r T_1^{-1} &= \bm t_{r+1}.
\end{align}
Note that we imposed the PBC on the tetrahedral chain.
The latter reads as
\begin{align}
    T_1\phi(r)T_1^{-1} &= \phi(r)+ \frac \pi 2, \\
    T_1 \theta(r) T_1^{-1} &= \theta(r) + \pi.
\end{align}
The lattice translation thus forbids $\cos(2\phi)$, $\sin(2\phi)$, $\cos\theta$, and $\sin \theta$.
Imposing the $\mathbb Z_3$ rotation and the lattice translation symmetries, we can forbid the relevant interactions except for $\partial_x\phi$ and $\partial_x\theta$.

Let us show that the time-reversal symmetry forbids them.
In the chirality-liquid phase I, the time-reversal~\eqref{TR_eff} works as the $\pi$ rotations around the $x$ axis at low energies [Eq.~\eqref{TR_pi_rot}].
$t_r^x=(t_r^++(t_r^+)^\dag)/2$ and $t_r^y=(t_r^+ - (t_r^-)^\dag)/2$ are written as
\begin{align}
    t_r^x &= (-1)^r b_0 \cos \theta +ib_1 \sin \theta \sin (2\phi), \\
    t_r^y &= (-1)^r b_0 \sin \theta -ib_1 \cos \theta \sin(2\phi),
\end{align}
because of $[\phi(x),\theta(x)]=i\pi/2$.
These bosonization formulas lead us to the following action of the time reversal $\mathcal T$:
\begin{align}
    \mathcal T \phi(r) \mathcal T^{-1} &= -\phi(r), \\
    \mathcal T \theta(r) \mathcal T^{-1} &= -\theta(r).
\end{align}
The $\mathbb Z_2^T$ symmetry thus forbids $\partial_x\phi$ and $\partial_x\theta$.
These interactions do not immediately open the gap but eventually does~\cite{Metlitski2018_anomaly}.
In fact, adding an interaction
\begin{align}
    g\sum_{r=1}^L \chi_r \approx 2g \sum_{r=1}^L t_r^z \approx \frac{2g}{\pi} \int_0^L dr \, \partial_r \phi,
\end{align}
to the Hamiltonian \eqref{H_eff} or \eqref{H_eff_bosonized} eventually opens the pseudospin gap by driving the chirality-liquid ground state to the spin-1 Haldane state.

\section{Summary}\label{sec:summary}

We discussed the gSPT phase of the geometrically frustrated tetrahedral spin chain \eqref{H_tetra} [Figs.~\ref{fig:tetrahedron}~(a) and (b)].
The unit cell contains one tetrahedron with four localized spins.
Within each tetrahedron, three bonds are antiferromagnetic ($J>0$) and  the other three bonds are ferromagnetic ($\alpha J<0$).
For $\alpha_c<\alpha<0$ with $\alpha_c\approx -0.27$, the ground state of this tetrahedral spin chain belongs to the gSPT phase [the phase I of Fig.~\ref{fig:tetrahedron}~(c)].

We amassed the numerical evidences that the phase I indeed qualifies as the gSPT phase (Sec.~\ref{sec:num}).
With the finite-size DMRG calculations under the OBC, we confirmed that the phase I is described by the $c=1$ CFT (Fig.~\ref{fig:EE_log_gSPT}) and at the same time, accompanied by the $S=1/2$ edge spin on each end of the spin chain (Fig.~\ref{fig:ES_gSPT}).

To get insight into the phase I, we further developed the degenerate perturbation theory (Sec.~\ref{sec:perturbation}).
The perturbation theory uncovered that the $c=1$ CFT is the TLL of the chirality liquid~\cite{Okunishi2012_spintube_chiral}.
The spin and chirality degrees of freedom are strongly coupled with each other at the Hamiltonian level \eqref{H_eff}.
Since the tetrahedral spin chain with four spins per unit cell can trivially open the spin gap, the low-energy physics is fully written in terms of the chirality.

In the last section \ref{sec:LSM}, we showed the ingappability of the chirality liquid under the $\mathbb Z_3\times \mathbb Z_2^T$ symmetry, the lattice translation symmetry, and the U(1) spin-rotation symmetry, where $\mathbb Z_3$ and $\mathbb Z_2^T$ refer to the $\mathbb Z_3$ spatial rotation around the legs and the time reversal, respectively.
Here, we showd the ingappability by extending the argument of Ref.~\cite{yao2021_LSM_STBC} based on the STBC.
This argument clarified the existence of the 't Hooft anomaly of the lattice system without mapping it to a quantum field theory.
The anomaly (or equivalently, the ingappability) is directly related to the stability of the gSPT phase.
Should the anomaly be absent, the ground state in the gSPT phase would easily acquire the excitation gap without spontaneously breaking any symmetry.
In our gSPT phase of the chirality liquid with $S=1/2$ edge spins, the $S=1/2$ edge states is protected by the same symmetry as that protects the spin-1 Haldane phase, namely, at least one of the $D_2\cong \mathbb Z_2\times\mathbb Z_2$ spin-rotation symmetry, the time-reversal symmetry, and the bond-centered inversion symmetry.

\begin{acknowledgements}
The authors are grateful to  Shigetoshi Sota, Yasuhiro Tada, and Hiroshi Ueda for fruitful discussions.
This work was supported by  JSPS Grants-in-Aid for Transformative Research Areas (A) ``Extreme Universe'' Nos. JP21H05191 and 21H05182  (S.C.F.) and
JSPS KAKENHI Grant Nos. JP20K03769 and 21K03465 (S.C.F.).
\end{acknowledgements}

\appendix 

\section{Unpeturbed ground states}\label{app:unpeturbed_GS}

We write down the unperturbed ground state $\ket{\frac{1}{2},1,m,\chi}$ of $h_r$ in terms of $\ket{T_r^z,\chi}_{123}\ket{m}_4$, where $\ket{T_r^z,\chi}_{123}$ is the eigenstate of the $z$ component $T_r^z$ of $\bm{T}_r$ and the chirality $\chi=\pm 1$ and $\ket{m}_4$ is the eigenstate of $S_{r,\text{tot}}^z=m$.
Since $\bm{T}_r$ and $\bm{S}_{r,4}$ are ferromagnetically coupled to form the $S=1$ spin, the unperturbed ground state $\ket{\frac{1}{2},1,m,\chi}$ are written as
\begin{subequations}
\begin{align}
    \ket{\tfrac{1}{2},1,1,\chi} &= \ket{\tfrac{1}{2},\chi}_{123}\ket{\tfrac{1}{2}}_4, \\
    \ket{\tfrac{1}{2},1,0,\chi} &=\frac{1}{\sqrt{2}}( \ket{\tfrac{1}{2},\chi}_{123}\ket{-\tfrac{1}{2}}_4+ \ket{-\tfrac{1}{2},\chi}_{123}\ket{\tfrac{1}{2}}_4), \\
    \ket{\tfrac{1}{2},1,-1,\chi} &= \ket{-\tfrac{1}{2},\chi}_{123}\ket{-\tfrac{1}{2}}_4.
\end{align}
\label{GS_tetrahedron}
\end{subequations}

\noindent
It is helpful to look into contents of $\ket{\pm\frac{1}{2},\chi}_{123}$.
Let $\ket{m_1m_2m_3}$ be an eigenstate of $S_{r,n}^z=m_n$ for $n=1,2,3$.
We represent the eigenvalue $m_n=1/2$ and $-1/2$ as $m_n=\uparrow$ and $\downarrow$, respectively.
For example, $\ket{\uparrow_1\uparrow_1\downarrow_3}$ be an eigenstate of $S_{r,n}^z$ for $n=1,2,3$ with eigenvalues $S_{r,1}^z=S_{r,2}^z=1/2$ and $S_{r,3}^z=-1/2$.
We can write $\ket{\pm\frac{1}{2},\chi}_{123}$ as 
\begin{subequations}
\begin{align}
    \ket{\tfrac 12,+}_{123}
    &= \frac{1}{\sqrt{3}} (
    \ket{\downarrow_1\uparrow_2\uparrow_3}+\omega \ket{\uparrow_1\downarrow_2\uparrow_3} + \omega^{-1}\ket{\uparrow_1\uparrow_2\downarrow_3}), \\
    \ket{\tfrac{1}{2},-}_{123}
    &= \frac{1}{\sqrt{3}} (
    \ket{\downarrow_1\uparrow_2\uparrow_3}+\omega^{-1} \ket{\uparrow_1\downarrow_2\uparrow_3} + \omega \ket{\uparrow_1\uparrow_2\downarrow_3}), \\
    \ket{-\tfrac{1}{2},+}_{123}
    &= \frac{1}{\sqrt{3}} (
    \ket{\uparrow_1\downarrow_2\downarrow_3}+\omega \ket{\downarrow_1\uparrow_2\downarrow_3} + \omega^{-1}\ket{\downarrow_1\downarrow_2\uparrow_3}), \\
    \ket{-\tfrac{1}{2},-}_{123}
    &= \frac{1}{\sqrt{3}} (
    \ket{\uparrow_1\downarrow_2\downarrow_3}+\omega^{-1} \ket{\downarrow_1\uparrow_2\downarrow_3} + \omega\ket{\downarrow_1\downarrow_2\uparrow_3}),
\end{align}
\end{subequations}

\noindent
with $\omega=\exp(2\pi i/3)$.
These eigenstates satisfy
\begin{align}
    \chi_r^z \ket{\pm\tfrac{1}{2},\chi} &= \chi \ket{\pm\tfrac{1}{2},\chi},
\end{align}
where $\chi_r$ is a scalar chirality of the base triangle \eqref{chirality_def}.

\section{Pseudospin representation}
\label{app:pseudospin}

Here, we show technical details of the pseudospin representation of $\bm{S}_{r,n}$.
Let us denote the projection operator onto the unperturbed ground state of $h_r$ as $P_r$.
We can explicitly write $P_r$ as 
\begin{align}
    P_r
    &=\sum_{m=1,0,-1}\sum_{\chi=\pm} \ket{\tfrac{1}{2},1,m,\chi}\bra{\tfrac{1}{2},1,m,\chi}.
\end{align}
We adopt a basis to represent $\ket{\tfrac{1}{2},1,m,\chi}$ as 
\begin{subequations}
\begin{align}
    \ket{\tfrac{1}{2},1,1,+} &=\ket{1}= (1,0,0,0,0,0)^\top, \\
    \ket{\tfrac{1}{2},1,0,+} &=\ket{2}= (0,1,0,0,0,0)^\top, \\
    \ket{\tfrac{1}{2},1,-1,+} &=\ket{3}= (0,0,1,0,0,0)^\top, \\
    \ket{\tfrac{1}{2},1,1,-} &=\ket{4}= (0,0,0,1,0,0)^\top, \\
    \ket{\tfrac{1}{2},1,0,-} &=\ket{5}= (0,0,0,0,1,0)^\top, \\
    \ket{\tfrac{1}{2},1,-1,-} &=\ket{6}= (0,0,0,0,0,1)^\top.
\end{align}
\end{subequations}

\noindent
With this basis, we can represent $\bm{T}_r$ as
\begin{align}
    P_rT^z_r P_r
    &= P_r
    \begin{pmatrix}
    1/2 & 0 & 0 & 0 & 0 & 0 \\
    0 & 0 & 0 & 0 & 0 & 0 \\
    0 & 0 & -1/2 & 0 & 0 & 0 \\
    0 & 0 & 0 & 1/2 & 0 & 0 \\
    0 & 0 & 0 & 0 & 0 & 0 \\
    0 & 0 & 0 & 0 & 0 & -1/2
    \end{pmatrix}
    P_r
    \notag \\
    &= P_r \biggl(\frac 12 \mathbbm{1}_2 \otimes s_r^z\biggr)P_r, \\
    P_r T_r^+ P_r
    &= P_r
    \begin{pmatrix}
    0 & -1/\sqrt{2} & 0 & 0 & 0 & 0 \\
    0 & 0 & -1/\sqrt{2} & 0 & 0 & 0 \\
    0 & 0 & 0 & 0 & 0 & 0 \\
    0 & 0 & 0 & 0 & -1/\sqrt{2} & 0 \\
    0 & 0 & 0 & 0 & 0 & -1/\sqrt{2} \\
    0 & 0 & 0 & 0 & 0 & 0
    \end{pmatrix}
    P_r
    \notag \\
    &= P_r\biggl(-\frac 12\mathbbm{1}_2\otimes s_r^+\biggr)P_r,
    \label{T+_proj} \\
    P_r T_r^- P_r
    &= P_r \biggl( -\frac 12 \mathbbm{1}_2\otimes s_r^-\biggr)P_r,
    \label{T-_proj}
\end{align}
where $\mathbbm{1}_n$ denotes the $n\times n$ identity matrix.
The total spin $\bm{T}_r$ of the base triangle thus reads as the spin-1 operator $\frac{1}{2}\bm{s}_r$ at low energies.
As expected, $\bm{T}_r$ acts as the identity $\mathbbm{1}_2$ on the chirality degrees of freedom.
The minus sign on the right hand side of Eqs.~\eqref{T+_proj} and \eqref{T-_proj} can be eliminated by a global symmetry operation transformation,
\begin{align}
    U_z:=\exp\biggl(i\pi\sum_{r=1}^L\sum_{n=1}^4 S_{r,n}^z\biggr).
\end{align}
The application of $U_z$ alters no results in this paper because we impose the U(1) spin-rotation symmetry around $S^z$ on the system.
$U_z$ is included in this U(1) group.
In what follows, we redefine $U_z\bm{S}_{r,n}U_z^{-1}=(-S_{r,n}^x,-S_{r,n}^y,S_{r,n}^z)$ as $\bm{S}_{r,n}$ for simplicity.
This redefinition enables us to simplify the projection of $\bm{T}_r$:
\begin{align}
    P_r \bm{T}_r P_r = P_r \biggl(\frac 12\mathbbm{1}_2\otimes\bm{s}_r\biggr)P_r
    \label{T_pseudospin}
\end{align}

Since two spin-1/2 $\bm{S}_{r,4}$ and $\bm{T}_r$ form the spin-1 $\bm{S}_{r,\text{tot}}$, it immediately follows that 
\begin{align}
    P_r\bm{S}_{r,4}P_r =P_r\bm{T}_rP_r = P_r\biggl( \frac 12\mathbbm{1}_2\otimes\bm{s}_r\biggr)P_r.
    \label{S4_proj}
\end{align}
The other spins $\bm{S}_{r,n}$ ($n=1,2,3$) nontrivially act on the chirality degrees of freedom.
For example,
\begin{widetext}
\begin{align}
    P_r S_{r,3}^z P_r
    &=
    P_r
    \begin{pmatrix}
    1/6 & 0 & 0 & -\omega^{-1}/3 & 0 & 0 \\
    0 & 0 & 0 & 0 & 0 & 0 \\
    0 & 0 & -1/6 & 0 & 0 & \omega^{-1}/3 \\
    -\omega/3 & 0 & 0 & 1/6 & 0 & 0 \\
    0 & 0 & 0 & 0 & 0 & 0 \\
    0 & 0 & \omega/3 & 0 & 0 & -1/6
    \end{pmatrix}
    P_r
    \notag \\
    &= P_r \biggl( \frac{1}{6}(\mathbbm{1}_2-2\omega^{-1} t_r^+-2\omega t_r^-)\otimes s_r^z \biggr)P_r, \\
    P_rS_{r,3}^+P_r
    &= P_r
    \begin{pmatrix}
    0 & 1/3\sqrt{2} & 0 & 0 & -\sqrt{2}\omega^{-1}/3 & 0 \\
    0 & 0 & 1/3\sqrt{2} & 0 & 0 & -\sqrt{2}\omega^{-1}/3 \\
    0 & 0 & 0 & 0 & 0 & 0 \\
    0 & -\sqrt{2}\omega/3 & 0 & 0 & 1/3\sqrt{2} & 0 \\
    0 & 0 & -\sqrt{2}\omega/3 & 0 & 0 & 1/3\sqrt{2} \\
    0 & 0 & 0 & 0 & 0 & 0
    \end{pmatrix}
    P_r
    \notag \\
    &=P_r \biggl(  \frac 16 (\mathbbm{1}_2-2\omega^{-1}t_r^+ -2\omega t_r^-)\otimes s_r^+
    \biggr)P_r.
\end{align}
Repeating the same procedure for $n=1$ and $2$, we obtain
\begin{align}
    P_r \bm{S}_{r,n} P_r &= P_r \biggl( \frac 16(\mathbbm{1}_2-2\omega^{n-1}t_r^+-2\omega^{-(n-1)}t_r^-)\otimes\bm{s}_r\biggr)P_r,
    \label{S123_proj}
\end{align}
for $n=1,2,3$.
This pseudospin representation of $\bm{S}_{r,n}$ reproduces that of $\bm{T}_r$ [Eq.~\eqref{T_pseudospin}].

Finally, we show how the time reversal $T$ and the $2\pi/3$ rotation $R_r$ looks like in the language of $\bm{s}_r$ and $\bm{t}_r$.
\begin{align}
     P_r \mathcal T  P_r
    &= P_r
    \begin{pmatrix}
    0 & 0 & 0 & 0 & 0 & 1 \\
    0 & 0 & 0 & 0 & 1 & 0 \\
    0 & 0 & 0 & 1 & 0 & 0 \\
    0 & 0 & 1 & 0 & 0 & 0 \\
    0 & 1 & 0 & 0 & 0 & 0 \\
    1 & 0 & 0 & 0 & 0 & 0
    \end{pmatrix}P_r
    \notag \\
    &= P_r \biggl( 2t_r^x \otimes [2(s_r^x)^2-\mathbbm{1}_3] \biggr)P_r
    \notag \\
    &= P_r \biggl(-2t_r^x \otimes \exp(i\pi s_r^x) \biggr) P_r,
    \label{TR_proj}
    \\
    P_r R_r P_r
    &= P_j
    \begin{pmatrix}
    \omega & 0 & 0 & 0 & 0 & 0 \\
    0 & \omega & 0 & 0 & 0 & 0 \\
    0 & 0 & \omega & 0 & 0 & 0 \\
    0 & 0 & 0 & \omega^{-1} & 0 & 0 \\
    0 & 0 & 0 & 0 & \omega^{-1} & 0 \\
    0 & 0 & 0 & 0 & 0 & \omega^{-1}
    \end{pmatrix}P_r
    \notag \\
    &= P_r \biggl( \exp\biggl(i\frac{4\pi}{3}t_r^z \biggr)\otimes\mathbbm{1}_3 \biggr)P_r.
    \label{Z3_proj}
\end{align}
\end{widetext}

\end{document}